Author for correspondence:
Vinay Kumar Gupta
e-mail: vinay.gupta@warwick.ac.uk

THE ROYAL SOCIETY PUBLISHING

# Coupled constitutive relations: a second law based higher-order closure for hydrodynamics

Anirudh Singh Rana[1,2], Vinay Kumar Gupta[2,3] and Henning Struchtrup[4]

[1]Institute of Advanced Study, University of Warwick, Coventry CV4 7HS, UK
[2]Mathematics Institute, University of Warwick, Coventry CV4 7AL, UK
[3]Department of Mathematics, SRM Institute of Science and Technology, Chennai 603203, India
[4]Department of Mechanical Engineering, University of Victoria, Victoria, British Columbia, Canada V8W 2Y2

ASR, 0000-0002-5241-7102; VKG, 0000-0001-5473-2256

In the classical framework, the Navier–Stokes–Fourier equations are obtained through the linear uncoupled thermodynamic force-flux relations which guarantee the non-negativity of the entropy production. However, the conventional thermodynamic description is only valid when the Knudsen number is sufficiently small. Here, it is shown that the range of validity of the Navier–Stokes–Fourier equations can be extended by incorporating the nonlinear coupling among the thermodynamic forces and fluxes. The resulting system of conservation laws closed with the coupled constitutive relations is able to describe many interesting rarefaction effects, such as Knudsen paradox, transpiration flows, thermal stress, heat flux without temperature gradients, etc., which cannot be predicted by the classical Navier–Stokes–Fourier equations. For this system of equations, a set of phenomenological boundary conditions, which respect the second law of thermodynamics, is also derived. Some of the benchmark problems in fluid mechanics are studied to show the applicability of the derived equations and boundary conditions.

## 1. Introduction

The classical Navier–Stokes–Fourier (NSF) equations are known to fail in describing small-scale flows, for which the Knudsen number—defined as the ratio of the molecular mean free path to a characteristic hydrodynamic length scale—is sufficiently large [1,2]. It is well established that the traditional NSF equations cannot describe strong non-equilibrium effects, which occur at high Knudsen numbers; for instance, the classical NSF equations are not able to describe the heat flux parallel to flow direction which is not forced by temperature gradient [3,4], non-uniform pressure profile and characteristic temperature dip in Poiseuille flow [5–7], non-Fourier heat flux in a lid-driven cavity where heat flows from low temperature to high temperature [8,9], etc.

Several approaches to irreversible thermodynamics are available to determine the properties of a system near equilibrium. Linear irreversible thermodynamics (LIT) [10,11] is based on the assumption of *local thermodynamic equilibrium*, where thermal and caloric state equation and the Gibbs equation locally retain the forms they have in equilibrium. The Gibbs equation and the conservation laws for mass, momentum and energy are then combined to derive a mathematical form of the second law of thermodynamics—the balance equation for entropy. The requirement of positivity of the entropy generation rate leads directly to the constitutive laws for stress tensor and heat flux, resulting in the well-known laws of NSF [12,13].

Rational thermodynamics (RT) [14], in an attempt to relax the requirement of local thermodynamic equilibrium, postulates a particular form of the balance law for entropy, the Clausius–Duhem equation. Here, the non-convective entropy flux is prescribed to be the heat flux divided by the thermodynamic temperature—a relation that is one of the results of LIT. A careful evaluation of the conservation laws together with the entropy equation results in constitutive equations for stress tensor and heat flux. For simple fluids, RT gives the same constitutive equations as LIT, including the relations that describe local thermodynamic equilibrium [13].

Hence, although their postulates differ, both approaches—LIT and RT—appear to be equivalent for simple fluids. A particular feature of both approaches is the form of the entropy generation rate as the sum of products of thermodynamic forces and thermodynamic fluxes. The forces describe deviation from global thermodynamic equilibrium, and typically are gradients, e.g. those of temperature and velocity. The fluxes, e.g. heat flux and stress tensor, describe processes that aim to reduce the forces. For processes that are not too far from equilibrium, as described by LIT and RT, forces and fluxes are mathematically uncoupled, in the sense that the fluxes do not appear in the expressions for the forces.

The increasing miniaturization of physical devices has directed attention to the strong non-equilibrium conditions where the classical equations derived by LIT and RT lose their validity, and must be enhanced by proper extensions of the methods of derivation. In the present contribution, considering rarefied gas flows, we present an enhancement, based on a correction of the entropy flux that is suggested from the Kinetic Theory of Gases and Extended Thermodynamics (ET) [15].

A detailed description of a gas flow, ranging from near equilibrium to strong non-equilibrium conditions, is offered by the Boltzmann equation, which solves for the microscopic distribution function of gas molecules [16]. However, being a nonlinear integro-differential equation, the Boltzmann equation is difficult to solve and its direct solutions are computationally expensive. An alternative, but complementary, modelling of a gas can be done through macroscopic description, in which the behaviour of a gas is described by moments of the distribution function [2,17]. The main aim of the macroscopic modelling is to reduce the complexity by considering the transport equations for a finite number of (low-dimensional) moments—referred to as moment equations—instead of solving the Boltzmann equation for the (high-dimensional) distribution function. It is worth to note that the physical quantities, such as density, temperature, velocity, stress tensor and heat flux, in a gas appear as moments of the distribution function. Moment equations form an open hierarchy of equations, thus requiring a suitable closure. In kinetic theory, there are many approximation methods for closing a set of moment equations [17]. The well-known approximation methods include the Hilbert expansion method [1], the Chapman–Enskog (CE) expansion method [18], the Grad's moment method [19], regularized moment method [20,21], and entropy maximization [22,23]. Another approximation method, which derives a system of algebraic nonlinear coupled constitutive relations (NCCR) using the so-called *balanced closure* for the conservation laws was proposed by Myong [24].

The CE expansion method relies on the smallness of the Knudsen number. At zeroth- and first-order approximations the method leads to the Euler and NSF equations, and is fully equivalent to the results of LIT and RT (local equilibrium, etc.). At the second- and third-order approximations the method yields the Burnett and super-Burnett equations, respectively, which exhibit instabilities for time-dependent problems [25], and are thermodynamically inconsistent [26,27]. During the last decade, several modified forms of the Burnett equations have been suggested in the literature (e.g. [28–31]) that are stable; however, at present no proper boundary conditions are available for any of these sets of equations, hence their applicability is limited. On the other hand, the linearized Grad's 13 (also the R13) moment equations comply with the second law of thermodynamics, which also leads to thermodynamically consistent boundary conditions for these set of equations [32]. However, no entropy law has been established for the nonlinear moment equations.

In the following, we propose a phenomenological procedure in which the entropy flux contains a nonlinear contribution in stress and heat flux, which is motivated by results from rational extended thermodynamics (RET) [15]. In contrast to RET, where, similar to the moment method, the set of variables is enlarged to contain non-equilibrium variables, our approach still uses the basic equilibrium variables and fulfills the main conditions of local thermodynamic equilibrium (classical thermal and caloric equations of state, Gibbs equation). However, the additional term in the entropy flux yields additional terms in the entropy generation, which, as will be seen, can still be written as a sum of products of forces and fluxes, but now the fluxes appear explicitly in the forces, i.e. the fluxes are coupled through the additional terms in the forces.

We shall show that the conservation laws together with the resulting coupled constitutive relations (CCR) can capture many interesting non-equilibrium effects, such as Knudsen paradox, transpiration flows, thermal stress, heat flux without temperature gradients, etc., in good agreement with experiments and with kinetic theory, e.g. the solution of the Boltzmann equation.

The remainder of the paper is structured as follows. The derivation of the CCR for the conservation laws is detailed in §2. A thermodynamically consistent set of boundary conditions complementing the system of the conservation laws closed with the CCR (referred to as the *CCR model* hereafter) is presented in §3. The linear stability of the CCR model is analysed in §4 to show that the CCR model is stable to small perturbations. Classical flow problems of Knudsen minimum, heat transfer in an isothermal lid-driven cavity and normal shock structure are investigated in §5. The paper ends with conclusion in §6.

## 2. Derivation of coupled constitutive relations

The conservation laws, which are evolution equations for mass density $\rho$, macroscopic velocity $v_i$ and temperature $T$, read

$$\frac{D\rho}{Dt} + \rho \frac{\partial v_k}{\partial x_k} = 0, \tag{2.1}$$

$$\rho \frac{Dv_i}{Dt} + \frac{\partial p}{\partial x_i} + \frac{\partial \Pi_{ik}}{\partial x_k} = \rho F_i, \tag{2.2}$$

$$\rho c_v \frac{DT}{Dt} + p \frac{\partial v_k}{\partial x_k} + \frac{\partial q_k}{\partial x_k} + \Pi_{kl} \frac{\partial v_k}{\partial x_l} = 0. \tag{2.3}$$

Here, D/Dt denotes the convective time derivative, $p$ is the pressure, $\Pi_{ik}$ is the viscous stress tensor, $q_k$ is the heat flux and $F_i$ is the external force per unit mass. Throughout the paper, Einstein summation is assumed over the repeated indices unless stated otherwise.

We shall consider monatomic ideal gases only, for which the pressure is $p = \rho RT$ with R being the gas constant and specific internal energy is $u = c_v T$ with the specific heat $c_v = 3R/2$. Moreover, for monatomic ideal gases, the stress tensor is symmetric and tracefree, i.e. $\Pi_{kk} = 0$ [2].

It should be noted that conservation laws (2.1)–(2.3) contain the stress tensor $\Pi_{ik}$ and heat flux $q_k$ as unknowns, hence constitutive equations are required, which link the these quantities to the variables $\rho$, $v_i$ and $T$.

The second law of thermodynamics states that the total entropy of an isolated system can never decrease over time [10,13], that is

$$\rho \frac{Ds}{Dt} + \frac{\partial \Psi_k}{\partial x_k} = \Sigma \geq 0, \quad (2.4)$$

where $s$ denotes the specific entropy, $\Psi_k$ is the non-convective entropy flux and $\Sigma$ is the non-negative entropy generation rate. An important aspect of a constitutive theory is to determine the appropriate relations among the properties $s$, $\Psi_k$, $\Sigma$ and the variables $\rho$, $v_i$, $T$ and their gradients so that the closed conservation laws guarantee the second law of thermodynamics.

## (a) Uncoupled constitutive relations: the NSF equations

In LIT, the constitutive relations for closing the system of conservation laws (2.1)–(2.3), are obtained such that the second law of thermodynamics is satisfied for all thermodynamic processes. For this, in LIT, local thermodynamic equilibrium is assumed [10,12], which implies the local validity of the Gibbs equation

$$T \, ds = du - \frac{p}{\rho^2} \, d\rho. \quad (2.5)$$

For convenience, we shall write temperature $T$ in energy units as $\theta = RT$, and dimensionless entropy as $\eta = s/R$, so that the Gibbs equation (2.5) for a monatomic ideal gas leads to

$$\frac{D\eta}{Dt} = \frac{3}{2\theta} \frac{D\theta}{Dt} - \frac{1}{\rho} \frac{D\rho}{Dt}. \quad (2.6)$$

Multiplying equation (2.6) with $\rho$, and replacing $D\rho/Dt$, and $D\theta/Dt$ using the mass balance equation (2.1) and the energy balance equation (2.3), one obtains the entropy balance equation for LIT

$$\rho \frac{D\eta}{Dt} + \frac{\partial}{\partial x_k}\left(\frac{q_k}{\theta}\right) = -\frac{\Pi_{kl}}{\theta} \frac{\partial v_k}{\partial x_l} - \frac{q_k}{\theta} \frac{\partial \ln \theta}{\partial x_k}. \quad (2.7)$$

Comparison of equation (2.7) with equation (2.4) gives the LIT expression for non-convective entropy flux as $\Psi_k = q_k/\theta$ and that for the entropy production term as

$$\Sigma = -\frac{1}{\theta}\left(\Pi_{kl} \frac{\partial v_{\langle k}}{\partial x_{l\rangle}} + \frac{q_k}{\theta} \frac{\partial \theta}{\partial x_k}\right). \quad (2.8)$$

The angular brackets around indices represent the symmetric and traceless part of a tensor, for example,

$$\frac{\partial v_{\langle i}}{\partial x_{j\rangle}} = \frac{1}{2}\left(\frac{\partial v_i}{\partial x_j} + \frac{\partial v_j}{\partial x_i}\right) - \frac{1}{3}\frac{\partial v_k}{\partial x_k}\delta_{ij}, \quad (2.9)$$

where $\delta_{ij}$ is the Kronecker delta tensor.

The entropy production (2.8) assumes a canonical form, i.e.

$$\Sigma = \sum_{\alpha} \mathcal{J}_\alpha \mathcal{X}_\alpha,$$

with the thermodynamic fluxes $\mathcal{J}_\alpha = \{\Pi_{kl}, q_k\}_\alpha$ and the thermodynamic forces $\mathcal{X}_\alpha = -(1/\theta)\{(\partial v_{\langle k}/\partial x_{l\rangle}), (1/\theta)(\partial \theta/\partial x_k)\}_\alpha$. The phenomenological closure of LIT demands a linear relationship between fluxes and forces of the form $\mathcal{J}_\alpha = \sum_\beta \mathcal{L}_{\alpha\beta} \mathcal{X}_\beta$, where the matrix of phenomenological coefficients depends only on equilibrium properties, $\mathcal{L}_{\alpha\beta}(\rho, \theta)$, and must be non-negative definite. For proper transformations between different observer frames, stress and heat flux must be Galilean invariant tensors [13], and it follows that only forces and fluxes of the same tensor type (scalars, vectors, 2-tensors, etc.) can be linked (Curie Principle [10]). Accordingly,

$$\Pi_{ij} = -2\mu \frac{\partial v_{\langle i}}{\partial x_{j\rangle}} \quad \text{and} \quad q_i = -\kappa \frac{\partial \theta}{\partial x_i}, \tag{2.10}$$

where $\mu$ and $\kappa$ are the coefficients of shear viscosity and thermal conductivity, respectively, where factors with $\theta$ are absorbed in the coefficients.

It is worth pointing out that for a monatomic ideal gas interacting with power potentials, the viscosity depends on temperature alone as

$$\mu = \mu_0 \left(\frac{\theta}{\theta_0}\right)^w, \tag{2.11}$$

where $\mu_0$ is the viscosity at a reference temperature $\theta_0$ and $w$ is referred to as the viscosity exponent [2,16]. Furthermore, the heat conductivity is proportional to viscosity, $\kappa = 5\mu/(2\,Pr)$, where $Pr \simeq 2/3$ denotes the Prandtl number [2].

Relations $(2.10)_1$ and $(2.10)_2$ are the Navier–Stokes law and Fourier's law, respectively, and we refer to them as the linear *uncoupled constitutive relations*—emanating from LIT. When relations (2.10) are substituted in conservation laws (2.1)–(2.3), they yield the well-known compressible NSF equations of hydrodynamics.

## (b) Coupled constitutive relations

As mentioned in the Introduction, LIT and RT yield identical results for simple fluids—such as ideal gases—but differ in their assumptions. Indeed, RT assumes the entropy flux $\Psi_k = q_k/\theta$ that is an outcome in LIT. In a recent paper [33], Paolucci & Paolucci considered entropy flux as function of the hydrodynamic variables and their gradients in a complete nonlinear representation, but found that only the classical term $\Psi_k = q_k/\theta$ is compatible with the second law of thermodynamics. Nevertheless, allowing higher-order contributions for stress and heat flux, they found additions which are fully nonlinear in the gradients. Extended Irreversible Thermodynamics [15,34] is similar to LIT, only that non-equilibrium variables, in particular stress and heat flux, are added, with additional contributions to the Gibbs equation, in an attempt to go beyond local thermodynamic equilibrium. RET [15] proceeds differently, but also adds non-equilibrium variables, and yields a non-equilibrium Gibbs equation. The consideration of local non-equilibrium gives additions to the LIT entropy flux, which appears as explicit function of the extended set of variables.

In theories of (ET) for 13 moments, stress tensor $\Pi_{kl}$ and heat flux $q_i$ are field variables [15,34], and the entropy flux is expressed through these. Using representation theorems for isotropic tensor functions [13] and dimensional analysis, we find the most general entropy flux expression for these variables as

$$\Psi_k = \gamma \frac{q_k}{\theta} - \alpha \frac{\Pi_{kl} q_l}{p\theta} + \beta \frac{\Pi_{kl} \Pi_{lm} q_m}{p^2 \theta}, \tag{2.12}$$

where the dimensionless coefficients $\gamma$, $\alpha$, $\beta$ depend on the dimensionless invariants $(\Pi^2)_{ll}/p^2$, $(\Pi^3)_{ll}/p^3$, $q^2/(p^2\theta)$ (note that the invariant $\Pi_{kk} = 0$) [13].

Presently, we are only interested in the leading correction to the classical entropy flux $q_k/\theta$. Considering $\Pi_{kl}$ and $q_k$ as small and of the same order, Taylor expansion to second order yields

$$\Psi_k = \frac{q_k}{\theta} - \alpha_0 \frac{\Pi_{kl} q_l}{p\theta}, \tag{2.13}$$

where the coefficient of the first term on the right-hand side was chosen such that the classical result is reproduced, and $\alpha_0$ is a constant. The form (2.13) of the entropy flux appears as a result in RET of 13 moments (with $\alpha_0 = 2/5$) [15].

Interestingly, the gradient of the additional term $\alpha_0(\Pi_{kl}q_l/p\theta)$ can be expanded such that it yields higher-order additions to the thermodynamic forces. Indeed, introduction of this additional contribution in equation (2.7) yields an extended form of the second law that reads

$$\rho \frac{D\eta}{Dt} + \frac{\partial}{\partial x_k}\left(\frac{q_k}{\theta} - \underline{\alpha_0 \frac{\Pi_{kl} q_l}{p\theta}}\right) = -\frac{\Pi_{kl}}{\theta}\left[\frac{\partial v_{\langle k}}{\partial x_{l\rangle}} + \underline{\frac{\alpha_0}{p}\left(\frac{\partial q_{\langle k}}{\partial x_{l\rangle}} - \alpha_1 q_{\langle k} \frac{\partial \ln \theta}{\partial x_{l\rangle}} - \alpha_2 q_{\langle k} \frac{\partial \ln p}{\partial x_{l\rangle}}\right)}\right]$$

$$- \frac{q_k}{\theta^2}\left[\frac{\partial \theta}{\partial x_k} + \underline{\frac{\alpha_0}{\rho}\left(\frac{\partial \Pi_{kl}}{\partial x_l} - \alpha_1^* \Pi_{kl} \frac{\partial \ln \theta}{\partial x_l} - \alpha_2^* \Pi_{kl} \frac{\partial \ln p}{\partial x_l}\right)}\right]. \quad (2.14)$$

The underlined terms on the left-hand side of equation (2.14) are equal to the underlined terms on the right-hand side of equation (2.14), where $\alpha_1$ and $\alpha_2$ are arbitrary numbers, and $\alpha_r^* = 1 - \alpha_r$, $r \in \{1, 2\}$. The coefficients $\alpha_r$ and $\alpha_r^*$ distribute the contributions to entropy generation between different force-flux pairs; their values will be obtained from comparison to results from kinetic theory. For $\alpha_0 = 0$, the underlined terms in equation (2.14) vanish, and equation (2.7) is recovered.

The right-hand side of equation (2.14) is the entropy generation rate, which can—again—be recognized as a sum of products of the fluxes $\Pi_{kl}$ and $q_k$ with generalized thermodynamic forces (in square brackets). The corresponding phenomenological equations that guarantee positivity of the entropy production read

$$\Pi_{kl} = -2\mu\left[\frac{\partial v_{\langle k}}{\partial x_{l\rangle}} + \frac{\alpha_0}{p}\left(\frac{\partial q_{\langle k}}{\partial x_{l\rangle}} - \alpha_1 q_{\langle k} \frac{\partial \ln \theta}{\partial x_{l\rangle}} - \alpha_2 q_{\langle k} \frac{\partial \ln p}{\partial x_{l\rangle}}\right)\right] \quad (2.15)$$

and

$$q_k = -\frac{5\mu}{2 Pr}\left[\frac{\partial \theta}{\partial x_k} + \frac{\alpha_0}{\rho}\left(\frac{\partial \Pi_{kl}}{\partial x_l} - \alpha_1^* \Pi_{kl} \frac{\partial \ln \theta}{\partial x_l} - \alpha_2^* \Pi_{kl} \frac{\partial \ln p}{\partial x_l}\right)\right]. \quad (2.16)$$

Here, the coefficients were chosen such that in the classical limit (i.e. when $\alpha_0 = 0$), the NSF equations are obtained. Since the fluxes $\Pi_{kl}$ and $q_k$ appear explicitly in the forces, we refer to relations (2.15) and (2.16) as the CCR for conservation laws (2.1)–(2.3). Conservation laws (2.1)–(2.3) along with CCR (2.15) and (2.16) constitute the *CCR model*.

The extended hydrodynamic models, e.g. the Grad 13-moment equations, contain additional contribution to the entropy density as well as to the fluxes. On the other hand, in the CCR model a second-order contribution to the entropy flux is considered, while the Gibbs equation, and hence the entropy density, remain unchanged. We emphasize that unlike the Grad 13-moment equations, the full Burnett equations cannot be obtained from the CCR model. Hence the CCR model stands somewhere between the NSF equations (first order in $Kn$) and the Burnett equations (second order in $Kn$).

With the extended entropy flux (2.13) and coupling force-flux relations, we find additional linear contributions to stress and heat flux, in contrast to [33] where all additional contributions are strictly nonlinear. The linear contributions are well known in kinetic theory, and are required to describe processes. For instance, the linear contribution $\partial q_{\langle k}/\partial x_{l\rangle}$ in (2.15) describes thermal stresses, where temperature gradients can induce flow [4]. Similarly, the linear contribution $\partial \Pi_{kl}/\partial x_l$ in (2.16) describes stress-induced heat flux, as discussed in §5b.

Moment theories with 13 and more moments, as well as theories of ET, describe these effects as well, by providing full balance laws with time derivatives for higher moments. By contrast, the CCR model does not contain time derivatives. From this, one will in particular expect differences between the CCR model and ET in the description of transient processes, but only small differences for slow and steady-state processes. The benefit of the CCR model compared to available equations from ET and moment method is its full thermodynamic structure without restriction.

### (c) Evaluation of the phenomenological coefficients

The CCR (2.15) and (2.16) contain the coefficients $\alpha_0$, $\alpha_1$ and $\alpha_2$, which, in principle, can be determined from experiments or theoretical scenarios. While the Burnett equations are unstable in transient processes, their coefficients are obtained from the Boltzmann equation and they describe higher-order effects in gases with some accuracy. Hence, we determine the coefficients in the CCR from comparison with the Burnett equations.

**Table 1.** Phenomenological coefficients for hard-sphere (HS) and Maxwell molecule (MM) gases.

| molecule type | $Pr$ | $\alpha_0$ | $\alpha_2$ | $\alpha_1$ |
|---|---|---|---|---|
| MM | 2/3 | 2/5 | 0 | 0 |
| HS | 0.661 | 0.3197 | −0.2816 | 0.4094 |

The Burnett equations are obtained from the CE expansion in the Knudsen number, which is proportional to the viscosity. The procedure can be easily applied to the constitutive relations (2.15) and (2.16) as follows: stress and heat flux are expanded in terms of the viscosity, so that

$$\left. \begin{array}{l} \Pi_{kl} = \mu \Pi_{kl}^{(1)} + \mu^2 \Pi_{kl}^{(2)} + \cdots \\ q_k = \mu q_k^{(1)} + \mu^2 q_k^{(2)} + \cdots \end{array} \right\} \quad (2.17)$$

and

Inserting this ansatz into equations (2.15) and (2.16), we find at first order the NSF laws

$$\Pi_{kl}^{(1)} = -2 \frac{\partial v_{\langle k}}{\partial x_{l\rangle}} = -2 S_{kl} \quad \text{and} \quad q_k^{(1)} = -\frac{5}{2 Pr} \frac{\partial \theta}{\partial x_k}, \quad (2.18)$$

and at second order

$$\Pi_{kl}^{(2)} = \frac{5\alpha_0}{Pr} \frac{1}{p} \left[ \frac{\partial^2 \theta}{\partial x_{\langle k} \partial x_{l\rangle}} - \alpha_2 \frac{\partial \theta}{\partial x_{\langle k}} \frac{\partial \ln p}{\partial x_{l\rangle}} + (w - \alpha_1) \frac{\partial \theta}{\partial x_{\langle k}} \frac{\partial \ln \theta}{\partial x_{l\rangle}} \right] \quad (2.19)$$

and

$$q_k^{(2)} = \frac{5\alpha_0}{Pr} \frac{1}{\rho} \left[ \frac{\partial S_{kl}}{\partial x_l} - \alpha_2^* S_{kl} \frac{\partial \ln p}{\partial x_l} + \left( w - \alpha_1^* + \frac{1}{\alpha_0} \right) S_{kl} \frac{\partial \ln \theta}{\partial x_l} \right], \quad (2.20)$$

where $w$ is the exponent in the expression of viscosity for power potentials, see equation (2.11). Comparison of (2.19) and (2.20) with the Burnett constitutive relations (eqns (4.47) and (4.48) of [2], respectively) gives

$$\left. \begin{array}{l} \alpha_0 = \dfrac{Pr}{5} \varpi_3, \quad -\dfrac{5\alpha_0}{Pr} \alpha_2 = \varpi_4, \quad \dfrac{5\alpha_0}{Pr}(w - \alpha_1) = \varpi_5 \\ \alpha_0 = \dfrac{Pr}{5} \theta_4, \quad -\dfrac{5\alpha_0}{Pr}(1 - \alpha_2) = \theta_3, \quad \dfrac{5\alpha_0}{Pr}\left(w - 1 + \alpha_1 + \dfrac{1}{\alpha_0}\right) = 3\theta_5, \end{array} \right\} \quad (2.21)$$

and

where $\varpi_i$ and $\theta_i$ are the Burnett coefficients. Relations (2.21) yield the unknown coefficients as

$$\alpha_0 = \frac{Pr}{5} \varpi_3, \quad \alpha_2 = -\frac{\varpi_4}{\varpi_3}, \quad \alpha_1 = w - \frac{\varpi_5}{\varpi_3} \quad \text{or} \quad \alpha_1 = 1 - w + 3\frac{\theta_5}{\varpi_3} - \frac{5}{Pr}\frac{1}{\varpi_3}.$$

We note that the coefficients $\alpha_0, \alpha_1, \alpha_2$ in the CCR can be fitted to the Burnett coefficients in agreement with the well-known relations between Burnett coefficients, $\theta_4 = \varpi_3$ and $\varpi_3 + \varpi_4 + \theta_3 = 0$ [2,35].

The Burnett coefficients depend upon the choice of intermolecular potential function appearing in the Boltzmann collision operator, values of these coefficients for inverse-power law potentials can be found, for example, in [2,36]. The values of the phenomenological coefficients $\alpha_0, \alpha_1$ and $\alpha_2$ for the hard-sphere (HS) and Maxwell molecule (MM) gases are given in table 1; for other power potentials, they can be computed from equations (2.21). We emphasize that we have performed the expansion (2.17) only to determine the coefficients $\alpha_0, \alpha_1, \alpha_2$, but will use the full CCR as given in (2.15) and (2.16).

We also note that, at least for Maxwell molecules, the Burnett equations can be obtained by CE-like expansion of the 13 moment system (based on ET or Grad, higher moment sets yield the same). Hence, in the sense of the CE expansion, the CCR and ET models agree in the transport coefficients for those terms that the CCR model contain, but the CCR model offers a reduced model, which has less (Burnett) contributions than the full 13 moment system. In the following sections, we show that the CCR model—just as Grad method and ET—provides linearly stable equations (§4), and describe some classical flow problems in sufficient agreement to detailed kinetic models (§5).

## 3. Wall boundary conditions

Just as the process of finding constitutive relations in the bulk, the development of wall boundary conditions is based on the second law. Specifically, one determines the entropy generation at the interface, and finds the boundary conditions as phenomenological laws that guarantee positivity of the entropy generation.

The entropy production rate at the boundary $\Sigma_w$ is given by the difference between the entropy fluxes into and out of the surface [10], i.e.

$$\Sigma_w = \left(\Psi_k - \frac{q_k^w}{\theta^w}\right) n_k \geq 0. \tag{3.1}$$

Here, $n_k$ is unit normal pointing from the boundary into the gas, $q_k^w$ denotes the heat flux in the wall at the interface, and $\theta^w$ denotes the temperature of the wall at the interface. Here, the wall is assumed to be a rigid Fourier heat conductor, with the entropy flux $q_k^w/\theta^w$ and $\Pi_{ik}^w = 0$.

At the interface, the total fluxes of mass, momentum and energy are continuous, due to conservation of these quantities [10,32,37,38],

$$\left.\begin{aligned} v_k n_k &= v_k^w n_k = 0, \\ (p\delta_{ik} + \Pi_{ik})n_k &= p^w n_i \\ (pv_k + \Pi_{ik}v_i + q_k)n_k &= (p^w v_k^w + q_k^w)n_k, \end{aligned}\right\} \tag{3.2}$$

and

where all quantities with superscript w refer to wall properties, and the others refer to the gas properties. To proceed, we combine entropy generation and continuity conditions by eliminating the heat flux in the wall $q_k^w$ and the pressure $p^w$, and find, after insertion of the entropy flux (2.13),

$$\Sigma_w = -\left[\frac{q_k}{\theta\theta^w}\mathcal{T} + \Pi_{ik}\left(\alpha_0 \frac{q_i}{p\theta} + \frac{\mathcal{V}_i}{\theta^w}\right)\right] n_k \geq 0. \tag{3.3}$$

Here, $\mathcal{V}_i = v_i - v_i^w$ is the slip velocity, with $\mathcal{V}_i n_i = 0$, and $\mathcal{T} = \theta - \theta^w$ is the temperature jump.

To write the entropy generation properly as sum of products of forces and fluxes, it is necessary to decompose the stress tensor and heat flux into their components in the normal and tangential directions as [32]

$$\Pi_{ij} = \Pi_{nn}(\tfrac{3}{2}n_i n_j - \tfrac{1}{2}\delta_{ij}) + \bar{\Pi}_{ni} n_j + \bar{\Pi}_{nj} n_i + \tilde{\Pi}_{ij}, \quad \text{and} \quad q_i = q_n n_i + \bar{q}_i, \tag{3.4}$$

where $q_n = q_k n_k$, $\Pi_{nn} = \Pi_{kl} n_k n_l$, and

$$\left.\begin{aligned} \bar{q}_i &= q_i - q_n n_i, \\ \bar{\Pi}_{ni} &= \Pi_{ik} n_k - \Pi_{nn} n_i, \\ \tilde{\Pi}_{ij} &= \Pi_{ij} - \Pi_{nn}(\tfrac{3}{2}n_i n_j - \tfrac{1}{2}\delta_{ij}) - \bar{\Pi}_{ni} n_j - \bar{\Pi}_{nj} n_i, \end{aligned}\right\} \tag{3.5}$$

and

such that $\bar{q}_k n_k = \bar{\Pi}_{nk} n_k = \tilde{\Pi}_{kk} = \tilde{\Pi}_{ij} n_j = \tilde{\Pi}_{ij} n_i = 0$, see appendix A for more details.

Substituting equations (3.4) and (3.5) into equation (3.3), the entropy generation can be written as a sum of two contributions:

$$\Sigma_w = -\frac{\bar{\Pi}_{ni}}{p\theta}(\mathcal{P}\mathcal{V}_i + \alpha_0 \bar{q}_i) - \frac{(q_n + \bar{\Pi}_{ni}\mathcal{V}_i)}{p\theta\theta^w}(\mathcal{P}\mathcal{T} + \alpha_0 \Pi_{nn}\theta).$$

**Table 2.** Velocity-slip coefficient $\eta_{VS}$ and the temperature-jump coefficient $\eta_{TJ}$ for hard-sphere (HS) and Maxwell molecule (MM) gases obtained through the linearized Boltzmann equation in [40–43].

| molecule type | $\eta_{VS}$ | $\eta_{TJ}$ |
|---|---|---|
| MM | 1.1366 [40] | 1.1621 [42] |
| HS | 1.1141 [41] | 1.1267 [43] |

where $\mathcal{P} = p - \alpha_0 \Pi_{nn}$. As in LIT of the bulk, the positivity of entropy production is ensured by phenomenological equations

$$\bar{\Pi}_{ni} = -\frac{\varsigma_1}{\sqrt{\theta}}(\mathcal{P}\mathcal{V}_i + \alpha_0 \bar{q}_i) \quad \text{and} \quad q_n + \bar{\Pi}_{ni}\mathcal{V}_i = -\frac{\varsigma_2}{\sqrt{\theta}}(\mathcal{P}\mathcal{T} + \alpha_0 \Pi_{nn}\theta). \tag{3.6}$$

Here, $\varsigma_1$ and $\varsigma_2$ are non-negative coefficients, which can be obtained either from experiments or from kinetic theory models. We determine $\varsigma_1$ and $\varsigma_2$ from the Maxwell accommodation model [39]. This model employs the accommodation coefficient $\chi$, which is defined such that a fraction $\chi$ of the molecules hitting the wall returns to the gas in equilibrium with the wall properties (wall Maxwellian), whereas the remaining fraction $(1-\chi)$ is specularly reflected. Comparison with boundary conditions from this model shows that these coefficients are related to the velocity-slip coefficient $\eta_{VS}$ and the temperature-jump coefficient $\eta_{TJ}$ as [32]

$$\varsigma_1 = \sqrt{\frac{2}{\pi}} \frac{\chi}{2-\chi} \frac{1}{\eta_{VS}} \quad \text{and} \quad \varsigma_2 = \sqrt{\frac{2}{\pi}} \frac{\chi}{2-\chi} \frac{2}{\eta_{TJ}}. \tag{3.7}$$

The values of the velocity-slip coefficient $\eta_{VS}$ and the temperature-jump coefficient $\eta_{TJ}$ as obtained from the linearized Boltzmann equation [40–43] are given in table 2.

Conditions (3.6) are the required, and thermodynamically consistent, boundary conditions for the CCR model. The first term in the brackets of boundary condition $(3.6)_1$ relates the shear stress $(\bar{\Pi}_{ni})$ to the tangential velocity slip $(\mathcal{V}_i)$ while the second term describes thermal transpiration—a flow induced by the tangential heat flux. It is evident from boundary condition $(3.6)_1$ that the NSF equations, for which $\alpha_0 = 0$, do not allow thermal transpiration within the framework of thermodynamically consistent boundary conditions. Boundary condition $(3.6)_2$ relates temperature jump $(\mathcal{T})$ and viscous heating $(\bar{\Pi}_{ni}\mathcal{V}_i)$ to the normal heat flux $q_n$ [44].

Several authors have suggested second-order boundary conditions for the Navier–Stokes equations, which are in a form, possibly with other values of coefficients, identical to (3.6) [45]. Our derivation suggests that in order for the boundary conditions to be compatible with the second law of thermodynamics, the NSF equations do not suffice, while they arise quite naturally from the CCR.

We note that a full theory of boundary conditions is still missing in the theory of ET. Indeed, since the transport equations in ET are hyperbolic, one has to study the characteristics of the system in order to determine how many, and which, boundary conditions are required. We are not aware of complete sets of boundary conditions for nonlinear ET at present. For moment equations and linearized version of ET, one can use ideas from kinetic theory to determine boundary conditions. The approach suggested by Grad [19] violates Onsager symmetry conditions, and ideas as those above can be used to adjust some coefficients in order to guarantee the proper symmetries [32].

## 4. Linear stability analysis

In this section, we analyse the stability of the CCR model ((2.1)–(2.3) along with (2.15) and (2.16)) to small perturbations.

## (a) Linear dimensionless equations

For linear stability analysis, the CCR model is linearized by introducing small perturbations in the field variables from their values in an equilibrium rest state. We write

$$\rho = \rho_0(1+\hat{\rho}), \quad v_i = \sqrt{\theta_0}\,\hat{v}_i, \quad \theta = \theta_0(1+\hat{\theta}), \quad \Pi_{ij} = \rho_0\theta_0\hat{\Pi}_{ij}, \quad q_i = \rho_0\theta_0\sqrt{\theta_0}\hat{q}_i, \quad (4.1)$$

where $\rho_0$ and $\theta_0$ are the values of $\rho$ and $\theta$ in the equilibrium while the remaining variables vanish in the equilibrium rest state; hats denote the dimensionless perturbations in the field variables from their values in the equilibrium rest state; and $\sqrt{\theta_0}$ is the scale for making the velocity dimensionless. These dimensionless perturbations are assumed to be much smaller than unity so that the linear analysis remains valid. Consequently, the pressure is linearized as $p = p_0(1+\hat{p})$ with $p_0 = \rho_0\theta_0$ and $\hat{p} \approx \hat{\rho} + \hat{\theta}$. Furthermore, a relevant length scale $L$ is introduced so that the dimensionless space variable is $\hat{x}_i = x_i/L$ and the dimensionless time is $\hat{t} = t\sqrt{\theta_0}/L$. Also, the external force is assumed to be small (of the order of small perturbations) in a linear analysis, and the dimensionless external force is given by $\hat{F}_i = F_i L/\theta_0$. Substituting the values of field variables from equations (4.1) in the CCR model, introducing the dimensionless space and time variables, and retaining only the linear terms in the perturbation variables, one obtains

$$\frac{\partial \rho}{\partial t} + \frac{\partial v_k}{\partial x_k} = 0, \quad (4.2)$$

$$\frac{\partial v_i}{\partial t} + \frac{\partial p}{\partial x_i} + \frac{\partial \Pi_{ik}}{\partial x_k} = F_i, \quad (4.3)$$

$$\frac{3}{2}\frac{\partial \theta}{\partial t} + \frac{\partial v_k}{\partial x_k} + \frac{\partial q_k}{\partial x_k} = 0, \quad (4.4)$$

with

$$\Pi_{kl} = -2\,Kn\left(\frac{\partial v_{\langle k}}{\partial x_{l\rangle}} + \alpha_0\frac{\partial q_{\langle k}}{\partial x_{l\rangle}}\right) \quad \text{and} \quad q_k = -\frac{5}{2}\frac{Kn}{Pr}\left(\frac{\partial \theta}{\partial x_k} + \alpha_0\frac{\partial \Pi_{kl}}{\partial x_l}\right). \quad (4.5)$$

In equations (4.2)–(4.5) and henceforth, the hats are dropped for better readability, and

$$Kn = \frac{\mu_0}{\rho_0\sqrt{\theta_0}L}, \quad (4.6)$$

is the Knudsen number with $\mu_0$ being the viscosity of the gas in equilibrium. Thus, all quantities in equations (4.2)–(4.5) and henceforth are dimensionless unless otherwise mentioned.

## (b) Plane harmonic waves

Now we consider a one-dimensional process (in the $x$-direction) without any external forces (i.e. $F_i = 0$) and assume a plane wave solution of the form

$$\boldsymbol{\psi} = \overset{\circ}{\boldsymbol{\psi}}\exp[\mathrm{i}(\omega t - kx)], \quad (4.7)$$

for equations (4.2)–(4.5), where $\boldsymbol{\psi} = \{\rho, v_x, \theta, \Pi_{xx}, q_x\}^\top$, $\overset{\circ}{\boldsymbol{\psi}}$ is a vector containing the complex amplitudes of the respective waves, $\omega$ is the (dimensionless) frequency and $k$ is the (dimensionless) wavenumber. Substitution of the plane wave solution (4.7) into equations (4.2)–(4.5) gives a system of algebraic equations $\mathcal{A}\boldsymbol{\psi} = \mathbf{0}$, where

$$\mathcal{A} = \begin{bmatrix} \mathrm{i}\omega & -\mathrm{i}k & 0 & 0 & 0 \\ -\mathrm{i}k & \mathrm{i}\omega & -\mathrm{i}k & -\mathrm{i}k & 0 \\ 0 & -\mathrm{i}k & \frac{3}{2}\mathrm{i}\omega & 0 & -\mathrm{i}k \\ 0 & -\frac{4}{3}Kn\,\mathrm{i}k & 0 & 1 & -\frac{4}{3}Kn\,\alpha_0\mathrm{i}k \\ 0 & 0 & -\frac{5}{2}\frac{Kn}{Pr}\mathrm{i}k & -\frac{5}{2}\frac{Kn}{Pr}\alpha_0\mathrm{i}k & 1 \end{bmatrix}.$$

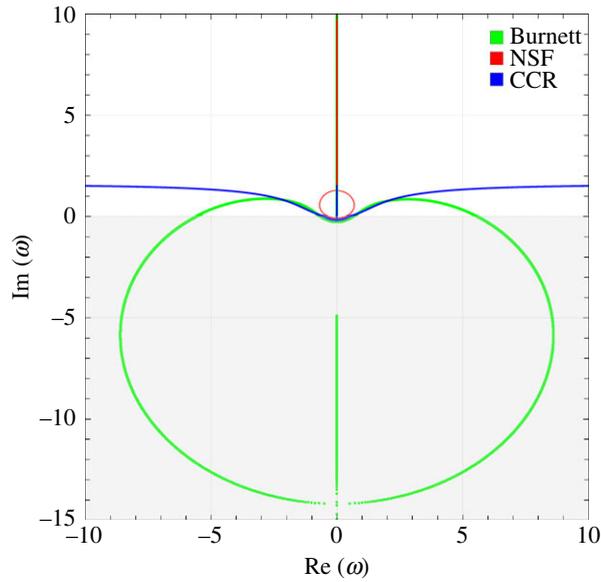

**Figure 1.** Temporal stability diagram for the CCR model (blue), NSF equations (red) and Burnett equations (green). The roots $\omega$ of the dispersion relations plotted in the complex plane for different values of $k$. The grey and white colours depict the unstable and stable regions, respectively. (Online version in colour.)

For non-trivial solutions, the determinant of matrix $\mathcal{A}$ must vanish, i.e. $\det(\mathcal{A}) = 0$. This leads to the dispersion relation, which is the relationship between $\omega$ and $k$:

$$\left(\frac{3}{2} + 5\frac{Kn^2}{Pr}\alpha_0^2 k^2\right)\omega^3 - i\left(2Kn + \frac{5}{2}\frac{Kn}{Pr}\right)k^2\omega^2 \\ - \frac{5}{2}\left[1 + \frac{2}{3}\frac{Kn^2}{Pr}(2 - 4\alpha_0 + 5\alpha_0^2)k^2\right]k^2\omega + \frac{5}{2}\frac{Kn}{Pr}ik^4 = 0. \quad (4.8)$$

For temporal stability, a disturbance is considered in space; consequently, the wavenumber $k$ is assumed to be real while the frequency $\omega$ can be complex. From the plane wave solution (4.7), it is clear that temporal stability requires the imaginary part of the frequency to be non-negative, i.e. $\text{Im}(\omega) \geq 0$, for all wavenumbers. In other words, if $\text{Im}(\omega) < 0$, then a small disturbance in space will blow up in time.

Figure 1 illustrates the stability diagram for the CCR model, NSF equations and Burnett equations shown by blue, red and green colours, respectively. The results for the NSF equations and Burnett equations are included for comparison. Assuming Maxwell molecules, we have $Pr = \frac{2}{3}$ and $\alpha_0 = \frac{2}{5}$; the Knudsen number is set to $Kn = 1$, that is the mean free path is used as relevant length scale. The figure exhibits $\omega(k) = \text{Re}(\omega) + i\,\text{Im}(\omega)$—obtained from the dispersion relations for the CCR model, NSF equations and Burnett equations—in the complex plane with $k$ as parameter. The grey region in the figure is the region, in which the stability conditions are not fulfilled, hence it depicts the unstable region, whereas the white region represents the stable one.

It is evident from figure 1 that the NSF equations (red) as well as the CCR model (blue) are linearly stable in time for all wavenumbers since the roots of their dispersion relations always have non-negative imaginary parts; on the other hand, the Burnett equations (green) are unstable in time, with negative roots for large wavenumbers. Grad-type moment systems and ET systems are linearly stable (e.g. [2,46]).

# 5. Classical flow problems

## (a) Knudsen minimum

The Knudsen minimum is observed in a force-driven Poiseuille flow of rarefied gases, in which, for given force, the mass flow rate of a gas first decreases with the Knudsen number, attains a minimum value around a critical Knudsen number and then increases with the Knudsen number. We shall investigate this problem through the CCR model.

Let us consider the steady state ($\partial(\cdot)/\partial t = 0$) of a gas confined between two isothermal, fully diffusive walls of a channel. Let the walls be located at (dimensionless positions) $y = \mp 1/2$ and be kept at a (dimensionless) temperature $\theta^w = 1$. The flow is assumed to be fully developed and driven by a uniform (dimensionless) external force $F$ in the positive $x$-direction parallel to the walls; all field variables are independent of $x$; and the velocity component in the $y$-direction is zero, i.e. $v_y = 0$. The problem can be described through the (linear-dimensionless) CCR model (equations (4.2)–(4.5)). For the problem under consideration, the mass balance equation (4.2) is identically satisfied and the rest of the equations simplify to

$$\frac{\partial \Pi_{xy}}{\partial y} = F, \quad \frac{\partial \rho}{\partial y} + \frac{\partial \theta}{\partial y} + \frac{\partial \Pi_{yy}}{\partial y} = 0, \quad \frac{\partial q_y}{\partial y} = 0, \tag{5.1}$$

with

$$\left.\begin{array}{l}\Pi_{xy} = -Kn\left(\dfrac{\partial v_x}{\partial y} + \alpha_0 \dfrac{\partial q_x}{\partial y}\right), \quad \Pi_{yy} = -\dfrac{4}{3} Kn\, \alpha_0 \dfrac{\partial q_y}{\partial y}, \\[6pt] q_x = -\dfrac{5}{2}\dfrac{Kn}{Pr}\alpha_0\dfrac{\partial \Pi_{xy}}{\partial y}, \quad q_y = -\dfrac{5}{2}\dfrac{Kn}{Pr}\left(\dfrac{\partial \theta}{\partial y} + \alpha_0\dfrac{\partial \Pi_{yy}}{\partial y}\right).\end{array}\right\} \tag{5.2}$$

and

Boundary conditions (3.6) at the walls (i.e. at $y = \mp\frac{1}{2}$) in the linear-dimensionless form reduce to

$$\pm \Pi_{xy} = -\varsigma_1(v_x + \alpha_0 q_x) \quad \text{and} \quad \pm q_y = -\varsigma_2 \alpha_0 \Pi_{yy}. \tag{5.3}$$

Solving equations $(5.1)_1$ and $(5.2)_{1,3}$ with boundary conditions $(5.3)_1$ yields a parabolic velocity profile

$$v_x = -\frac{1}{2}\frac{1}{Kn}F\left(y^2 - \frac{1}{4} - \frac{Kn}{\varsigma_1} - 5\frac{Kn^2}{Pr}\alpha_0^2\right).$$

Consequently, the mass flow rate of the gas is

$$\frac{1}{\sqrt{2}}\int_{-1/2}^{1/2} v_x \, dy = \frac{1}{2\sqrt{2}}F\left(\frac{1}{6}\frac{1}{Kn} + \frac{1}{\varsigma_1} + 5\frac{Kn}{Pr}\alpha_0^2\right). \tag{5.4}$$

Here, the additional factor of $1/\sqrt{2}$ in the mass flow rate is included in order to compare the present results with those obtained from the linearized Boltzmann equation (LBE) in [5] where the authors scale velocity by $\sqrt{2\theta}$. As the walls of the channel are fully diffusive, the accommodation coefficient $\chi$ is unity, and hence from equation $(3.7)_1$, $\varsigma_1 = (\sqrt{2/\pi})/\eta_{VS}$. The mass flow rate from the NSF equations with slip boundary conditions is obtained by setting $\alpha_0 = 0$, which gives $(1/2\sqrt{2})F((1/6)(1/Kn) + (1/\varsigma_1))$, while for NSF with no-slip boundary conditions one finds $(1/12\sqrt{2})(1/Kn)F$.

Figure 2 shows the mass flow rate of a hard-sphere gas plotted over the Knudsen number $\hat{Kn} = 4\sqrt{2}\,Kn/5$ for $F = 1$, as obtained from the CCR model (blue solid line), from the NSF equations with (red dashed line) and without (magenta solid line) slip boundary conditions, and from the LBE (symbols). Again, the Knudsen number $Kn$ is multiplied with a factor of $4\sqrt{2}/5$ in order to compare the results with those from [5]. The parameters $Pr$ and $\eta_{VS}$ for hard-spheres are taken from tables 1 and 2, respectively. It is clear from the figure that the CCR model predicts the Knudsen minimum and closely follow the mass flow rate profile from the LBE up to $Kn \simeq 1$. On the other hand, the NSF equations with or without slip do not predict the Knudsen minimum at all; in particular, the mass flow rate from the NSF equations with no-slip boundary conditions is not even close to that from the LBE equations. The mass flow rate from the CCR model (equation (5.4)) possesses the term $5(Kn/Pr)\alpha_0^2$, which dominates for large Knudsen numbers resulting into an increasing mass flow rate profile for large Knudsen numbers. However, such a term is not present in the expressions of mass flow rates obtained with the NSF equations with or without slip; consequently, the mass flow rate profiles from the NSF equations always decay with the Knudsen number.

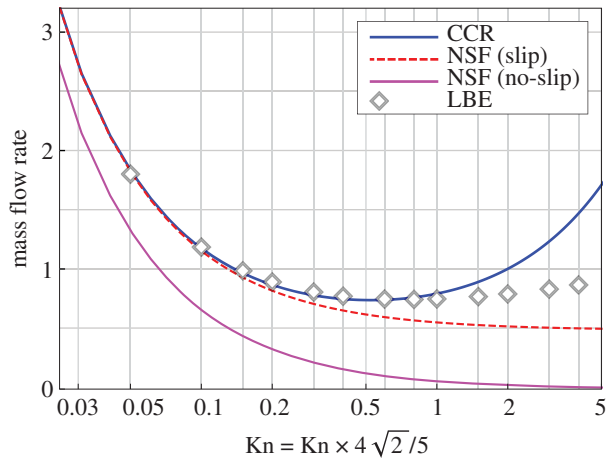

**Figure 2.** Mass flow rate of a hard-sphere gas in a force-driven Poiseuille flow plotted over the Knudsen number for $F=1$. The accommodation coefficient is $\chi=1$. (Online version in colour.)

### (b) Lid-driven cavity flow

The lid-driven cavity is a well-known test problem in rarefied gas dynamics, in which a gas—under no external force—is confined to a square enclosure of (dimensionless) length 1. The boundaries at $x=0$, $x=1$ and $y=0$ are stationary while the upper boundary at $y=1$ is moving in the $x$-direction with a velocity $v_{lid}$. The boundaries are kept at a constant temperature, which is equal to the initial temperature of the gas inside the cavity so that the dimensionless temperature of the walls $\theta^w=1$. The flow in the cavity is assumed to be in steady state and independent of the $z$-direction, i.e. $\partial(\cdot)/\partial t=0$ and $\partial(\cdot)/\partial z=0$. The problem is solved through the CCR model ((2.1)–(2.3) along with (2.15) and (2.16) in dimensionless form) numerically using a finite difference scheme whose details can be found in [9].

Figure 3 shows the vertical and horizontal components of the velocity along the horizontal and vertical centrelines of the cavity, respectively, computed through the CCR model and the NSF equations for $Kn=0.1/\sqrt{2}$ and $v_{lid}=0.21$ (50 m s$^{-1}$ in dimensional units). The results are compared to DSMC simulations for hard spheres, hence we chose the phenomenological coefficients for hard spheres from table 1. The velocity profiles from the CCR model as well as from the NSF equations are in good agreement with the DSMC simulations [8].

It is well known that the classical NSF laws cannot describe heat transfer phenomena in a lid-driven cavity (see e.g. [9,47,48] and references therein). Therefore, either extended models, such as Grad-type moment equations or the R13 equations [2,17], or other advanced constitutive laws ought to be used for the closure in order to describe the heat transfer characteristics.

We study in particular the temperature field and heat flux induced in the lid-driven cavity and compare the results from the CCR model to those from DSMC presented in [8]. Figure 4 illustrates the heat flux lines plotted over the temperature contours, and compares the predictions of (from left to right) the CCR model, DSMC and the NSF equations. The NSF equations predict that the heat flows from hot (top-right corner) to cold (top-left corner) in an orthogonal direction to the temperature contours. However, both the CCR model and DSMC predict heat flux from cold region to hot region, which is non-Fourier effect.

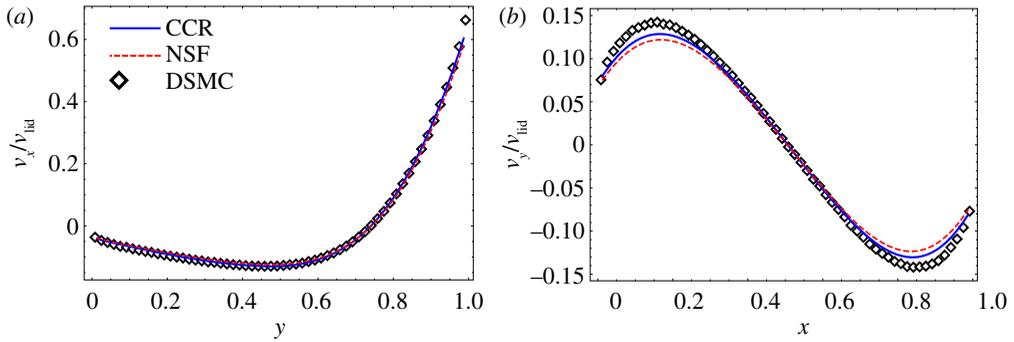

**Figure 3.** Dimensionless velocity profiles: (*a*) *x*-component of the velocity ($v_x/v_{\text{lid}}$) along the vertical centreline of the cavity, and (*b*) *y*-component of the velocity ($v_y/v_{\text{lid}}$) along the horizontal centreline of the cavity. Numerical solutions from the CCR model (blue solid lines) and from the NSF equations (red dashed lines) for $Kn = 0.1/\sqrt{2}$ are compared to the DSMC data (symbols) from [8]. (Online version in colour.)

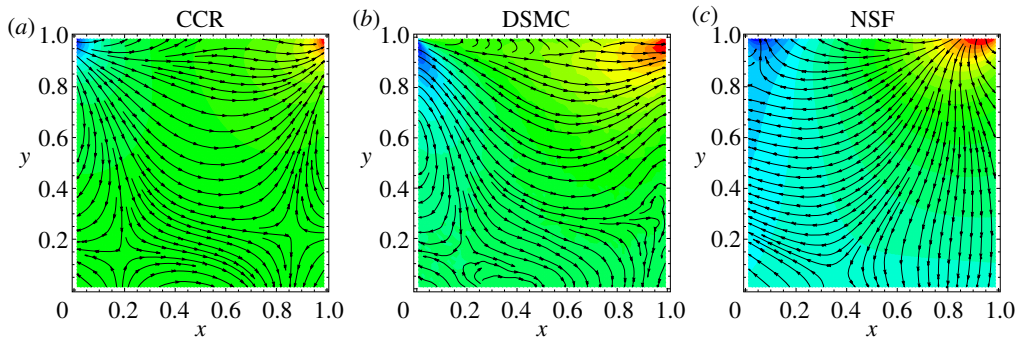

**Figure 4.** Heat flux lines superimposed over temperature contours for the lid-driven cavity problem at $Kn = 0.1/\sqrt{2}$ obtained through the CCR model (*a*) and the NSF equations (*c*) are compared to DSMC data (*b*) given in [8]. (Online version in colour.)

The non-Fourier heat flux can easily be understood from the linear terms itself, which dominate the nonlinear terms. From the (linearized) momentum balance equation (4.3) (ignoring the time derivative and external force terms), the divergence of the stress is given by

$$\frac{\partial \Pi_{ik}}{\partial x_k} = -\frac{\partial p}{\partial x_i}. \quad (5.5)$$

Substitution of this in the (linearized) constitutive relation for the heat flux $(4.5)_2$ yields

$$q_i \simeq Q_i = -\frac{5}{2}\frac{Kn}{Pr}\frac{\partial \theta}{\partial x_i} + \frac{5}{2}\frac{Kn}{Pr}\alpha_0 \frac{\partial p}{\partial x_i}. \quad (5.6)$$

The first term on the right-hand side of equation (5.6) is the Fourier's contribution to the heat flux while the second term is the non-Fourier contribution to the heat flux due to the pressure gradient, which is responsible for heat transfer from the cold region to the hot one. Figure 5 displays heat flux lines (black) and $Q_i$ lines (red) superimposed on the contours of $(\theta - \alpha_0 p)$ for the CCR model (*a*) and DSMC (*b*). It is evident from the figure that the heat flux $q_i$ from (2.16) is estimated well with $Q_i$ (equation (5.6)), which is orthogonal to $(\theta - \alpha_0 p)$ contour lines. It can also be seen by comparing figures 4 and 5 that the heat flux lines in DSMC simulations are governed by the $(\theta - \alpha_0 p)$ gradients, not by the temperature alone. Near the bottom of the cavity, the heat flux lines (black) given by DSMC differ from those given by $Q_i$ due to statistical noise inherent to the DSMC method.

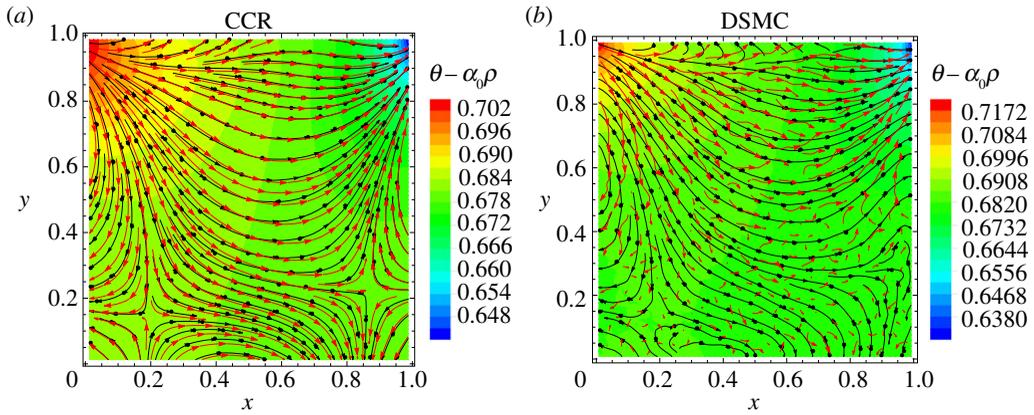

**Figure 5.** Heat flux lines (black) and $Q_i$ lines (red) superimposed over temperature contours for the lid-driven cavity problem at $Kn = 0.1/\sqrt{2}$ obtained through the CCR model (*a*) and DSMC data (*b*) from [8]. (Online version in colour.)

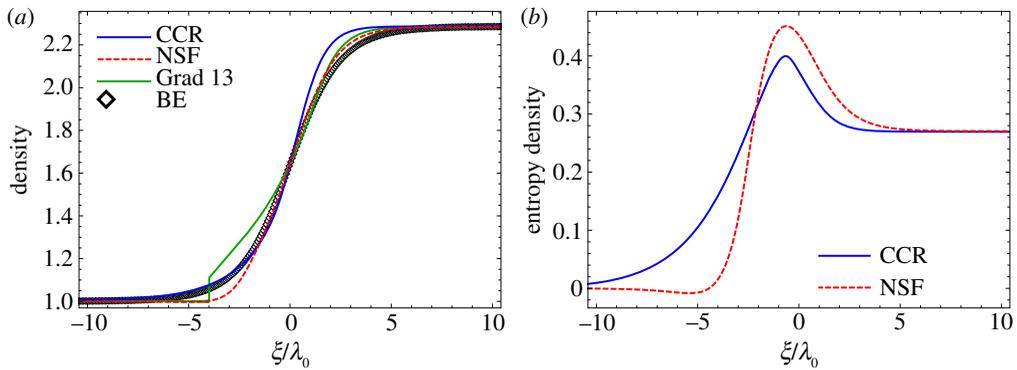

**Figure 6.** Shock structures for Ma = 2 plotted over $\xi/\lambda_0$: (*a*) density and (*b*) entropy density. Numerical solutions from the CCR model (blue solid lines), from the NSF equations (red dashed lines) and from the Grad 13-moment equations (green solid line) for $Kn = 1$ are compared to the DSMC data (symbols) from [46]. (Online version in colour.)

## (c) Normal shock structure

For one-dimensional normal shock structure, one considers a reference frame moving with the shock and hence all the quantities depend only on a single variable $\xi = x - v_x t$ [49,50]. The dimensionless hydrodynamic variables (denoted with bars) in upstream ($\xi \to -\infty$) are $\bar{\rho}_u = 1$, $\bar{v}_u = \sqrt{5/3}\, Ma$, $\bar{\theta}_u = 1$, where $\bar{v}_0$ is the dimensionless velocity in the $x$-direction and $Ma$ is the Mach number. Furthermore, the dimensionless hydrodynamic variables in downstream ($\xi \to \infty$), $\bar{\rho}_d$, $\bar{v}_d$ and $\bar{\theta}_d$, follow from the Rankine–Hugoniot conditions [46,49]:

$$\bar{\rho}_d = \frac{4\,Ma^2}{Ma^2+3}, \quad \bar{v}_d = \sqrt{\frac{5}{3}}\,\frac{Ma^2+3}{4\,Ma} \quad \text{and} \quad \bar{\theta}_d = \frac{(5Ma^2-1)(Ma^2+3)}{16\,Ma^2}. \tag{5.7}$$

The shock profiles are obtained by solving the different models (CCR, NSF and Grad) numerically using a central difference scheme employed in [46]. Figure 6 illustrates the profiles for (*a*) density and (*b*) entropy density for $Ma = 2$, $w = 1$ and $Kn = 1$ compared with the DSMC data given in [46]. In order to make $\xi$ dimensionless, the mean free path of the upstream region $\lambda_0 = 0.0014$ m [2,46] has been used. Both the CCR (blue solid line) and NSF (red dashed line) models give smooth shock profiles while the Grad 13-moment equations (green solid line) give sub-shock, which is not observed in the DSMC simulations (symbols). The sub-shock in Grad 13-moment equations is due to the hyperbolicity of the equations and occurs above the characteristic speed [46,49]. Clearly, both the CCR and NSF models mismatch the DSMC results. Nevertheless, the density profile from the CCR model agrees well with the DSMC results in the upstream part of the shock while being overpredicted in the downstream of the shock whereas that from the NSF equations has clear mismatch from the DSMC simulations both in the upstream and downstream parts of the shock. Notwithstanding, it is worthwhile to note that one of the main assumption made in the CCR model is the validity of local equilibrium, which implies that the entropy density $\eta$ depends only on the equilibrium variables and is given by the Gibbs equation (2.5), i.e.

$$\eta = \frac{3}{2} \ln \frac{\theta}{\theta_0} - \ln \frac{\rho}{\rho_0}. \tag{5.8}$$

This assumption may not be valid in strong non-equilibrium processes, such as problems involving shocks. Figure 6b exhibits the entropy density, which turns out to be non-monotonous for both the CCR (blue solid line) and NSF (red dashed line) models. Here we want to emphasize that while entropy is not monotonous, it certainly has positive production. Although, we could not find the Boltzmann/DSMC solution for the entropy density in the literature, it has been reported, for instance in [50], that the entropy density computed from the Boltzmann equation is monotonous. This again asserts that the local equilibrium assumption for entropy may not be valid for strongly non-equilibrium processes, and non-equilibrium variables must also be included in the expression of entropy (e.g. [33,50]) along with the entropy flux.

## 6. Conclusion

Combining ideas of different approaches to Irreversible Thermodynamics, in particular LIT, RT and RET, we derived an improved set of constitutive relations for stress tensor and heat flux— the CCR. The model describes processes in mildly rarefied gases in sufficient approximation, and reproduces important rarefaction effects, such as the Knudsen minimum, and non-Fourier heat transfer, which cannot be described by classical hydrodynamics (NSF).

By construction, the resulting transport equations are accompanied by a proper entropy inequality with non-negative entropy generation for all processes and are linearly stable. This is a clear distinction to other models for rarefied gases, such as the Burnett equations, which are unstable due to lack of a proper entropy, or Grad-type moment equations, which are accompanied by proper entropy inequalities, and are stable, only in the linear case [15].

Thermodynamically consistent boundary conditions for the CCR model have been developed as well, which describe velocity slip, temperature jump and transpiration flow at the boundaries.

The CCR add several higher order terms to the NSF system, in bulk and at the boundary. The CCR model is accompanied by an entropy inequality, in which the entropy remains the equilibrium entropy as integrated from the equilibrium Gibbs equation, but entropy flux and entropy generation exhibit higher order correction terms. The model gives a good description of some important rarefaction effects, but does not provide as fine resolution as the full Boltzmann equation, or higher order moment equations (e.g. the R13 and R26 equations [17,21]). In particular, Knudsen layers are not resolved, and only appear indirectly in the corrected jump and slip coefficients.

The development of macroscopic transport equations for rarefied gases at larger orders in the Knudsen number with a full formulation of the second law of thermodynamics is an important project within the field of non-equilibrium thermodynamics. Our results for the normal shock problem show that the assumption of local equilibrium may not be legitimate since problems involving shocks are intrinsically in strong non-equilibrium. A possible way to resolve this issue is to consider an extended form of entropy—involving non-equilibrium variables—along with non-equilibrium contribution to the entropy flux, which is planned for future work. On the other hand, one will accept an approximation of the second law, if the system of equations provides sufficient accuracy for the description of processes. For instance, the R13 equations provide good accuracy, but have a proven second law only in the linearized case. While this guarantees linear stability, little can be said about the nonlinear behaviour. On the other hand, cases where the full second law—linear and nonlinear—is enforced, are either not amenable to analytic closure [23], or not sufficiently accurate [51].

In steady state, the linearized CCR model reduces to the linearized Grad's 13-moment equations, which have been studied extensively in the literature. In particular, Green's functions solutions were obtained for the steady-state linearized Grad equations by Lockerby & Collyer [52]. The numerical framework based upon these fundamental solutions can readily be implemented for the linearized CCR model allowing for three-dimensional steady computation at remarkably low computational cost. The appropriate numerical recipes for the nonlinear CCR model would be challenging—especially due to the non-local coupling of stress and heat flux—which will be the subject of future research.

The successful development of thermodynamically consistent transport equations for rarefied gases at higher orders will only be possible by using the best of several approaches to irreversible thermodynamics and new ideas. We hope that the development of the CCR system based on ideas of LIT, RT and RET will be a useful step towards this end. We also envisage that these and similar ideas will prove useful for further development of recent moment models for monatomic gas mixtures [53,54] and granular gases [55].

Data accessibility. The research material (source code and data) can be accessed in the electronic supplementary material.
Authors' contributions. A.S.R. and H.S. initiated the work. A.S.R. and V.K.G. carried out the computations. All authors discussed the results, contributed to the writing of the manuscript and gave their final approval for publication.
Competing interests. We declare that we have no competing interests.
Funding. This work has been financially supported in the UK by EPSRC grant no. EP/N016602/1. A.S.R. gratefully acknowledges the funding from the European Union's Horizon 2020 research and innovation programme under the Marie Skłodowska Curie grant agreement no. 713548. V.K.G. gratefully acknowledges the financial support through the Commonwealth Rutherford Fellowship. H.S. gratefully acknowledges support through an NSERC Discovery grant.
Acknowledgements. The authors thank Dr James Sprittles and Prof. Duncan Lockerby for many fruitful discussions.

## Appendix A. Decomposition of a vector and a tensor

A vector $a_i$ and a symmetric traceless tensor $A_{ij}$ can be decomposed into their components in the direction of a given normal $\mathbf{n}$ and in the directions perpendicular to it (tangential directions) as follows. The vector $a_i$ is decomposed as

$$a_i = a_n n_i + \bar{a}_i,$$

where $a_n n_i$ with $a_n = a_k n_k$ is the normal component of $a_i$ while $\bar{a}_i = a_i - a_n n_i$ is the tangential component of $a_i$. By definition, $\bar{a}_i$ is such that $\bar{a}_i n_i = 0$.

The symmetric traceless tensor $A_{ij}$ is decomposed as [32]

$$A_{ij} = \tfrac{3}{2} A_{nn} n_{\langle i} n_{j\rangle} + \bar{A}_{ni} n_j + \bar{A}_{nj} n_i + \tilde{A}_{ij}, \tag{A 1}$$

where Einstein summation never applies to the indices '$n$' and $A_{nn} = A_{kl} n_k n_l$. The first term in the summation of the above decomposition is the component of the tensor $A_{ij}$ in the direction of normal; the second and third terms in the summation of the above decomposition are the normal-tangential components of $A_{ij}$; the fourth term in the summation of the above decomposition denotes the tangential-tangential component of $A_{ij}$. By definition, $\bar{A}_{nk} n_k = 0$ and $\tilde{A}_{ij} n_i = \tilde{A}_{ij} n_j = 0$. Multiplication of equation (A 1) with $n_j$ yields

$$\bar{A}_{ni} = A_{ij} n_j - A_{nn} n_i, \tag{A 2}$$

and equation (A 1) itself gives

$$\tilde{A}_{ij} = A_{ij} - \tfrac{3}{2}A_{nn}n_{\langle i}n_{j\rangle} - \bar{A}_{ni}n_j - \bar{A}_{nj}n_i. \tag{A 3}$$